\documentclass[preprint]{aastex}

\shorttitle{Formation of a Massive Black Hole in M82}
\shortauthors{Matsushita et al.}

\begin{document}

\slugcomment{ApJ Letter, accepted}

\title{Formation of a Massive Black Hole\\
	at the Center of the Superbubble in M82}

\author{Satoki Matsushita\altaffilmark{1},
\email{smatsushita@cfa.harvard.edu}
	Ryohei Kawabe\altaffilmark{2}, Hironori Matsumoto\altaffilmark{3},
	Takeshi G. Tsuru\altaffilmark{4}, Kotaro Kohno\altaffilmark{2},
	Koh-Ichiro Morita\altaffilmark{2}, Sachiko K. Okumura\altaffilmark{2},
	\and Baltasar Vila-Vilar\'o\altaffilmark{5}}

\altaffiltext{1}{Submillimeter Array, Harvard-Smithsonian
	Center for Astrophysics, P.O.\ Box 824, Hilo, HI 96721-0824}
\altaffiltext{2}{Nobeyama Radio Observatory,
	Minamimaki, Minamisaku, Nagano 384-1305, Japan}
\altaffiltext{3}{Center for Space Research, Massachusetts Institute of
	Technology, 77 Massachusetts Avenue, Cambridge, MA 02139-4307}
\altaffiltext{4}{Department of Physics, Faculty of Science,
	Kyoto University, Sakyo-ku, Kyoto, 606-8502, Japan}
\altaffiltext{5}{Steward Observatory, University of Arizona, Tucson, AZ 85721}

\begin{abstract}
We performed $^{12}$CO(1 -- 0), $^{13}$CO(1 -- 0), and HCN(1 -- 0)
interferometric observations of the central region (about 450 pc in radius)
of M82 with the Nobeyama Millimeter Array, and
have successfully imaged a molecular superbubble and spurs.
The center of the superbubble is clearly shifted from the nucleus by 140 pc.
This position is close to that of the massive black hole (BH) of
$\gtrsim460$ M$_{\odot}$ and the $2.2\mu$m secondary peak
(a luminous supergiant dominated cluster),
which strongly suggests that these objects may be related to
the formation of the superbubble.
Consideration of star formation in the cluster based on the infrared data
indicates that (1) energy release from supernovae can account for
the kinetic energy of the superbubble, (2) the total mass of stellar-mass BHs
available for building-up the massive BH may be much higher than 460 M$_{\odot}$,
and (3) it is possible to form the middle-mass BH of $10^{2}-10^{3}$ M$_{\odot}$
within the timescale of the superbubble.
We suggest that the massive BH was produced and is growing in the intense
starburst region.
\end{abstract}

\keywords{black hole physics, galaxies: individual (M82, NGC 3034),
	galaxies: ISM, galaxies: starburst, ISM: bubbles}

\section{INTRODUCTION}
\label{intro}

A starburst is a star formation event with a high star formation efficiency.
This usually occurs in the galactic nuclear/central regions.
Starbursts and successive supernova explosions cause drastic changes in
kinematic and physical conditions of interstellar medium,
and produce (expanding) superbubbles, chimneys, and/or large-scale outflows.
There is observational evidence for these structures in hot and/or atomic gas
in many galaxies \citep[e.g.][]{fab89,ten88}, but there is much less clear
evidence in molecular gas.
Starburst phenomena in the nuclear regions are also thought to be closely
related to active galactic nuclei (AGNs) powered by massive black holes
(MBHs; more massive than a stellar-mass BH),
such as fueling AGNs and/or causing the growth of MBHs.
One question, ``can starbursts produce the seeds of MBHs or
middle-mass BHs?,'' is still largely unresolved.

Nearby \citep[3.25 Mpc;][]{san75} irregular galaxy M82 (NGC 3034) has
kilo-parsec scale bipolar outflows which can be seen by optical emission
lines \citep[e.g.][]{sho98} and X-ray \citep[e.g.][]{bre95}
(Fig.~\ref{bubble}a).
This outflow is believed to be made by frequent supernova explosions
at the central region in consequence of starburst.
Recent X-ray studies have found evidences of an MBH with a mass of
$\gtrsim460 M_{\odot}$ in the starburst region \citep{mat99,pta99}.
In this region, however, strong dust absorption prevents detailed
optical/near-infrared observations, so we performed millimeter-wave
observations which would not be affected by the dust.

\section{OBSERVATIONS}
\label{obs}

Aperture synthesis observations of the central region of M82 were carried out
in the $^{12}$CO, $^{13}$CO, and HCN $J=1\rightarrow0$ lines
(rest frequency = 115.271 GHz, 110.201 GHz, and 88.632 GHz, respectively)
with the Nobeyama Millimeter Array (NMA) during 1997 November - 1999 March.
All the images were obtained using three configurations of six 10m antennas
which are equipped with tunerless SIS receivers \citep{sun94},
and the system noise temperatures in the single side band were about 950 K,
850 K, and 450 K at 115 GHz, 110 GHz, and 89 GHz, respectively.
As a back-end, we used an XF-type spectro-correlator Ultra Wide Band
Correlator \citep[UWBC;][]{oku00}, with a total bandwidth of 512 MHz over
256 channels for the $^{12}$CO observations (corresponding to 1300 km s$^{-1}$
bandwidth with 5.2 km s$^{-1}$ velocity resolution), and with a total
bandwidth of 1024 MHz over 128 channels for the $^{13}$CO and HCN observations
(corresponding to 2800 km s$^{-1}$ and 3500 km s$^{-1}$ bandwidths with
22 km s$^{-1}$ and 27 km s$^{-1}$ velocity resolutions, respectively).
The band-pass calibration was done with 3C273, and 0923+392 was observed
every 10 minutes as a phase and amplitude calibrator.
The flux scale of 0923+392 was determined by comparisons with Mars and Uranus,
and has an uncertainty of $\sim20\%$.

\section{EXPANDING MOLECULAR SUPERBUBBLE}
\label{exp-mol-bub}

The overall distribution in the $^{12}$CO total integrated intensity map
shows diffuse spurs north and south (minor axis direction) of the galaxy,
in addition to the previously identified \citep{she95} three prominent
peaks (the so-called North-East [NE] and South-West [SW] lobes and
a central peak).
These spurs trace the dark filaments which can be seen in optical B-band
images \citep{alt99} (Fig.~\ref{bubble}b).
Similar structures have been detected further out in the filaments
\citep{nak87,kun97,alt99} which are connected to the large-scale H$\alpha$
and X-ray outflow \citep{sho98,bre95} (Fig.~\ref{bubble}a),
but ours is the first detection close to the disk.

A position-velocity (PV) diagram of $^{12}$CO data cut along the major axis
of the galaxy (Fig.~\ref{pv}b) shows an arc-like deviation from rigid
rotation between the central peak and the SW lobe.
Such deviations are common features of expanding \ion{H}{1} superbubbles
\citep{deu90}.
In addition, recent literature also suggests that this deviation implies
the existence of a molecular superbubble \citep{nei98,wei99,wil99}.
However, any shell structures have not been detected in spatial image
so far, and \citet{wei99} concluded that the superbubble had already
broken toward the minor axis direction of this galaxy.
We made a channel map, binning over the velocity range $V_{\rm LSR}=118$
to 212 km s$^{-1}$, which corresponds to the range deviates from the rigid
rotation (Fig.~\ref{bubble}c).
The map clearly shows a shell-like structure, with a diameter of
$\sim210$ pc $\times$ 140 pc ($\sim14''\times9''$)
elongated toward north-south direction, between the two peaks.
This structure is also visible in other molecular lines such as $^{13}$CO
(Fig.~\ref{bubble}d) and HCN (Fig.~\ref{bubble}e).
These images indicate that the superbubble is still not broken out of
the galactic disk.
Around the superbubble, however, there are some spurs connecting with
the shell structure, so that it may be possible that some parts of
the superbubble were already broken.

If we assume that the shell structure is the result of edge-brightening
\citep{wil99}, the southern part of the shell should be moving perpendicular
to the line-of-sight with nearly the rigid rotation velocity,
which corresponds to the systemic velocity of the molecular superbubble.
Indeed, the velocity field of the southern shell is very close to
rigid rotation (Fig.~\ref{pv}c).
In addition, the PV diagram shows that almost all of the molecular gas
except that of the superbubble is on the rigid rotation (Fig.~\ref{pv}b).
These reasons lead us to conclude that the expansion velocity of
the superbubble is the largest velocity deviation from the rigid rotation
seen in the PV diagram;
the resultant expansion velocity is $\sim100$ km s$^{-1}$.
However, we cannot reject the possibility of an expansion velocity of
$\sim50$ km s$^{-1}$, which is the mean velocity deviation in the PV
diagram \citep{wei99}, so we use the velocity range of $50-100$ km s$^{-1}$
for the following calculations.
Using the velocity and the size of the superbubble derived above, the elapsed
time from the explosion can be estimated as $\sim(1-2)\times10^{6}$ years.
\citet{wei99} also derived similar timescale of $1\times10^{6}$ years.

Next, we will estimate the energetics of the superbubble.
Using the intensity ratios between the $^{12}$CO line and the $^{13}$CO and
HCN lines, we calculated the CO-to-H$_{2}$ conversion factor based on
the Large-Velocity-Gradient (LVG) approximation \citep{sak99}.
The resultant conversion factor is $(1.4\pm0.6)\times10^{20}$
cm$^{-2}$ (K km s$^{-1}$)$^{-1}$,
which is consistent with the previous observations of
$\sim(1.0-1.2)\times10^{20}$ cm$^{-2}$ (K km s$^{-1}$)$^{-1}$
\citep{wil92,smi91}.
Hence, we adopted our estimated value, and the calculated molecular
superbubble mass would be $\sim1.8\times10^{8}$ M$_{\odot}$.
From this molecular gas mass and the expansion velocity, we derive
the kinetic energy of the superbubble to be $\sim(0.5-2)\times10^{55}$ erg,
equivalent to the total energy of $\sim10^{3}-10^{4}$ supernovae,
and an order of magnitude or more larger than that observed in
\ion{H}{1} superbubbles in other galaxies
\citep[$<10^{54}$ erg;][]{ten88}.
The energy estimated by \citet{wei99} was $\sim2\times10^{54}$ erg,
similar to ours, although their estimation have a large ambiguity
in the volume of the material swept up by the explosion.

\section{MASSIVE BLACK HOLE INSIDE THE SUPERBUBBLE}
\label{mbh-in-bub}

At the center of the superbubble, there is a hard X-ray variable point source
and a 2.2 $\mu$m secondary peak (Fig.~\ref{bubble}c);
hard X-ray observations with ASCA indicate that there is a strong point
source located in the central region of M82.
This source shows time variability in its intensity,
which indicates the existence of a MBH with its mass of
$\sim460-2\times10^{8}$ M$_{\odot}$ \citep{mat99,pta99}.
Comparing the ROSAT HRI image \citep{bre95,ste99} with the false ROSAT
image made from ASCA data (the image using similar energy-band with that of
ROSAT), it was found that the position of the hard X-ray variable source is
consistent with that of the X-ray peak detected with the ROSAT \citep{mat99}.
Detailed comparison between the location of this point source
and our newly obtained $^{12}$CO map, reveals that the X-ray point source is
located at the center of the superbubble.
Recent high resolution X-ray observations with Chandra show
that there is a variable source inside the superbubble with its peak
luminosity of $\sim9\times10^{40}$ erg s$^{-1}$, assuming this source has
a same spectrum as the ASCA variable source \citep{mat00}.
This result also supports the conclusion that there is a MBH inside
the superbubble.
On the other hand, the emission from the 2.2 $\mu$m secondary peak
\citep{die86,les90} seems to be dominated by luminous supergiants
\citep{joy87}, which suggest that it is a late phase dense starburst
cluster.

Since these objects are located close to the center of the superbubble,
it is natural to think that these objects are related to the superbubble's
formation.
We therefore discuss the possibility that the starburst at
the 2.2 $\mu$m secondary peak produces the superbubble and the MBH.
We first calculated the stellar population of this 2.2 $\mu$m secondary peak
cluster assuming an initial mass function (IMF), and estimated
the starburst evolution as follows:
The luminosity of the 2.2 $\mu$m secondary peak with an extent (full width
at half maximum) of $4''$ \citep{mcl93} is equivalent to $\sim1500$ M2
supergiants.
We assume that stars with initial mass larger than 30 M$_{\odot}$
\citep[short-lived stars with lifetimes $<2\times10^{6}$ yr;][]{lan98}
have already exploded as supernovae, and that the remaining stars with
initial masses of $25-30$ M$_{\odot}$ are now M2 supergiants.
Using an extended Millar-Scalo IMF \citep{ken83} of $dN/dm\propto m^{-2.5}$
with lower and upper mass limits of 1 and 100, and assuming that there are
1500 stars whose masses are $25-30$ M$_{\odot}$, the total stellar numbers
and the total mass formed at the 2.2 $\mu$m secondary peak cluster would be
about $8\times10^{5}$ stars and $2\times10^{6}$ M$_{\odot}$, respectively.
The number of $\ge30$ M$_{\odot}$ stars which would have already exploded
in this cluster is $\sim4\times10^{3}$, consistent with the estimated
number of supernovae needed to create the superbubble ($10^{3}-10^{4}$).
The observational evidence and IMF calculations suggest that the expanding
molecular superbubble may be a result of localized starbursts
which occurred around the position of the 2.2 $\mu$m secondary peak.

We next discuss the possibility of making a MBH at the starburst in M82.
If we assume that stars with initial mass of $>25$ M$_{\odot}$ would
create stellar-mass BHs of almost the same mass \citep{bro94},
there would be $\sim4\times10^{3}$ BHs with a total mass of
$\sim2\times10^{5}$ M$_{\odot}$ in the  2.2 $\mu$m secondary peak cluster.
Assuming an isothermal sphere stellar density distribution \citep{lee95},
about $4\times10^{3}$ M$_{\odot}$ BHs would sink into the cluster center
by dynamical friction within $2\times10^{6}$ yr.
This mass is well within the range of that of the MBH estimated from
the hard X-ray observations.
There is also a possibility of star-star merger.
If we assume a merging probability of 0.1\% in the cluster, which is
a similar probability to the simulations of stellar mergers in star
clusters \citep{por99}, it is possible to make one $\sim700$ M$_{\odot}$
star, or several stars of a few hundred M$_{\odot}$, in the cluster.
Explosions of such very high mass stars can produce $\gtrsim10^{52}$ erg or
even $10^{54}$ erg energy hypernovae \citep[e.g.][]{pac98}, which can be seen
as $\gamma$-ray bursts, and may possibly have created the superbubble and MBH.
If we set the lower limit of the BH mass, $M_{\rm BH}$, in this
cluster using the hard X-ray observations,
and the upper limit using the total mass of the stars which seem to be
already exploded ($\ge30$ M$_{\odot}$),
the range of the BH mass would be $460-2\times10^{5}$ M$_{\odot}$.
This mass range suggests that this MBH might be a middle-mass BH.
Since the cluster mass, $M_{\rm cluster}$, is calculated as $2\times10^{6}$
M$_{\odot}$ using the IMF, the ratio of $M_{\rm BH}$ to $M_{\rm cluster}$
would be $-3.6\lesssim\log(M_{\rm BH}/M_{\rm cluster})\lesssim-1.0$.
As shown in Figure~\ref{bh-ratio}, this range is on
the $M_{\rm bulge}$-$M_{\rm BH}$ relation for galaxies with
supermassive BHs \citep{mag98}.

Since there are many other clusters in the central region of M82,
another question, ``why other clusters do not have MBHs?,'' would arise.
One possibility is lack of stellar density.
If their densities are small, the effect of dynamical friction or
star-star merging rate may decrease
and it may not be possible to make massive objects.
The low density clusters would also be affected by tidal force
\citep[e.g.][]{tan00} and tend to be smaller clusters which are not
large enough to make massive objects.
There is another possibility that the MBH was not created at the cluster
but supplied from satellite galaxy by minor merger \citep[e.g.][]{tan00}.
In this case, there are two possibilities of the agreement between
the position of the MBH and that of the superbubble; one is just accidentally
overlapping each other, and another is that the superbubble is created by
the MBH.
A BH merging with a compact object (neutron star, white dwarf, etc.)
can cause a large energy release \citep[$\sim10^{54}$ erg;][]{mes99}.
If the MBH meets with a cluster on the way sinking toward the galactic center,
there are possibilities to merge with some compact objects,
and as a result, energetic explosions would occur and the superbubble would
be made.
These discussions still have large ambiguities because of lack of detailed
information, so that further observations with many
wavelengths/frequencies would be needed.

In future, this middle-mass BH may increase its mass.
Since massive stars still exist ($\sim10^{5}$ M$_{\odot}$) at the 2.2 $\mu$m
secondary peak cluster, it is possible to feed stars and/or stellar-mass BHs
to the middle-mass BH.
Also, the middle-mass BH is still not at the dynamical center of M82,
and it may sink down to the center with dynamical friction, where mass is
strongly concentrated.
Therefore there are many ways for the middle-mass BH in M82 to increase in
mass and therefore grow up to the supermassive BH.
Our results may give a new explanation to the reason why
some of the quasars are embedded in interacting
(and therefore active star forming) galaxies, 
and strong $\gamma$-ray bursts and middle-mass BHs are located either
at or offset from galactic nuclei.

\acknowledgements

We would like to thank J. Makino, Y. Funato, and anonymous referee
for helpful discussion and comments.
We also thank B. Wallace for carefully reading our manuscript.
We are grateful to the NRO staff for the operation and improvement of NMA.
S.M. and H.M. are financially supported by the JSPS
Postdoctoral Fellowships for Research Abroad.

\clearpage

\clearpage

\figcaption{Various scale images of M82.
	In molecular gas images, value for the colorscale
	is indicated on the top of each figure,
	and the plus mark indicates the position of the galactic
	nucleus determined from the peak of the strongest $2.2\mu$m source
	\citep{les90}.
	(a) The X-ray image \citep[contours;][]{bre95} overlaid on
	the H$\alpha$ image \citep[grayscale;][]{sho98}.
	(b) $^{12}$CO integrated intensity map (contours)
	overlaid on optical B-band image \citep[grayscale;][]{alt99}.
	Bright regions on the grayscale are indicated in black and dark regions
	are in white.
	The contour levels of the $^{12}$CO map are
	$5, 10, 15, 20, 25, 30, 40, 50, \cdots, 90\sigma$,
	where 1 $\sigma$ = 1.64 Jy beam$^{-1}$ km s$^{-1}$ [= 23 K km s$^{-1}$].
	The synthesized beam (2\farcs8 $\times$ 2\farcs3
	or 42 pc $\times$ 35 pc) is shown at the bottom-left corner.
	(c) The $^{12}$CO molecular superbubble image, binning over the velocity
	range of $118-212$ km s$^{-1}$.
	The contour levels of the $^{12}$CO map are $-10, 10, 20, 30, \cdots,
	110\sigma$,
	where 1 $\sigma$ = 8.9 mJy beam$^{-1}$ [= 127 mK].
	The synthesized beam size is the same as (b).
	The diamond mark and the filled circle indicate the central positions
	of the $2.2\mu$m secondary peak \citep{die86}, and the X-ray point
	source observed with ROSAT \citep{ste99} which corresponds with
	the ASCA hard X-ray variable source position \citep{mat99}.
	We also indicate the recently observed Chandra X-ray variable source
	\citep{mat00} with a circle, which radius corresponds to its position
	uncertainty.
	(d) The HCN molecular superbubble image, binning over the velocity
	range of $134-215$ km s$^{-1}$.
	The contour levels of the HCN map are $-6, -3, 3, 6, 9, 12, 15\sigma$,
	where 1 $\sigma$ = 3.7 mJy beam$^{-1}$ [= 39 mK].
	The synthesized beam (4\farcs1 $\times$ 3\farcs6
	or 62 pc $\times$ 54 pc) is shown at the bottom-left corner.
	(e) The $^{13}$CO molecular superbubble image, binning over the velocity
	range of $127-214$ km s$^{-1}$.
	The contour levels of the $^{13}$CO map are
	$-3, 3, 6, 9, 12, 15, 18\sigma$,
	where 1 $\sigma$ = 5.4 mJy beam$^{-1}$ [= 41 mK].
	The synthesized beam (3\farcs9 $\times$ 3\farcs4
	or 59 pc $\times$ 51 pc) is shown at the bottom-left corner.
\label{bubble}}

\figcaption{Position-velocity (PV) diagrams of M82.
	Values for the colorscale are indicated on the top of each figure.
	(a) $^{12}$CO integrated intensity map.
	Solid lines indicate the sliced regions for PV diagrams displayed below.
	The plus mark is the same as Fig.~\ref{bubble}b.
	(b) PV diagram at slice A.
	The solid line indicates rigid rotation velocity.
	The region of the superbubble
	(around $\alpha$(B1950)=$9^{\rm h}51^{\rm m}43^{\rm s}.6$)
	clearly deviates from the rigid rotation velocity.
	(c) PV diagram at slice B.
	The solid line indicates rigid rotation velocity.
	Almost all of the gas at this slice is on the rigid rotation velocity.
\label{pv}}

\figcaption{Correlation diagram between the mass of black holes ($M_{\rm BH}$)
	and those of the host bulges or clusters ($M_{\rm bulge}$) in galaxies.
	The bar mark indicates the estimated mass range of the middle-mass BH 
	in M82, assuming its host corresponds to the 2.2 $\mu$m secondary peak
	cluster.
	The upper limit of the middle-mass BH mass and the 2.2 $\mu$m
	secondary peak cluster mass have been estimated from IMF calculations,
	and the lower limit of the BH mass has been estimated from the hard 
	X-ray observations.
	Cross marks indicate the data taken from \citet{mag98}, and a solid
	line is a linear fitting of their data.
	We did not include their upper limit data in this diagram.
	This figure clearly shows that the data of M82 is on the trend of
	the mass of the supermassive black holes and those of the host bulges.
\label{bh-ratio}}


\begin{thebibliography}{99}
\bibitem[Alton et al.(1999)]{alt99} Alton, P. B., Davies, J. J.,
	\& Bianchi, S.  1999, \aap, 343, 51
\bibitem[Bregman et al.(1995)]{bre95} Bregman, J. N., Schulman, E.,
	\& Tomisaka, K.  1995, \apj, 439, 155
\bibitem[Brown \& Bethe(1994)]{bro94} Brown, G. E., \& Bethe, H. A. 1994,
	\apj, 423, 659
\bibitem[Deul \& den Hartog(1990)]{deu90} Deul, E. R.,
	\& den Hartog, R. H.  1990, \aap, 229, 362
\bibitem[Dietz et al.(1986)]{die86} Dietz, R. D., Smith, J., Hackwell, J. A.,
	Gehrz, R. D., \& Grasdalen, G. L. 1986, \aj, 91, 758
\bibitem[Fabbiano(1989)]{fab89} Fabbiano, G.  1989, \araa, 27, 87
\bibitem[Joy et al.(1987)]{joy87} Joy, M., Lester, D. F.,
	\& Harvey, P. M.  1987, \apj, 319, 314
\bibitem[Kennicutt(1983)]{ken83} Kennicutt, R. C.  1983, \apj, 272, 54
\bibitem[Kuno \& Matsuo(1997)]{kun97} Kuno, N., \& Matsuo, H. 1997,
	\pasj, 49, 265
\bibitem[Lang(1998)]{lan98} Lang, K. R.  1998, Astrophysical Formulae,
	(Heidelberg: Springer-Verlag)
\bibitem[Lee(1995)]{lee95} Lee, H. M.  1995, \mnras, 272, 605
\bibitem[Lester et al.(1990)]{les90} Lester D. F., Carr, J. S., Joy, M.,
	\& Gaffney, N.  1990, \apj, 352, 544
\bibitem[Magorrian et al.(1998)]{mag98} Magorrian, J., et al.  1998,
	\aj, 115, 2285
\bibitem[Matsumoto \& Tsuru(1999)]{mat99} Matsumoto, H., \& Tsuru, T. G.  1999,
	\pasj, 51, 321
\bibitem[Matsumoto et al.(2000)]{mat00} Matsumoto, H., et al.  2000,
        \apjl \ Letter, in press (astro-ph/0009250)
\bibitem[McLeod et al.(1993)]{mcl93} McLeod, K. K., Rieke, G. H., Rieke, M. J.,
	\& Kelly, D. M.  1993, \apj, 412, 111
\bibitem[M\'esz\'aros et al.(1999)]{mes99} M\'esz\'aros, P., Rees, M.J.,
	\& Wijers, R.A.M.J.  1999, New Astronomy, 4, 303
\bibitem[Nakai et al.(1987)]{nak87} Nakai, N., Hayashi, M., Handa, T.,
	Sofue, Y., \& Hasegawa, T.  1987, \pasj, 39, 685
\bibitem[Neininger et al.(1998)]{nei98} Neininger, N., Gu\'elin, M.,
	Klein, U., Garc\'ia-Burillo, S., \& Wielebinski, R.  1998,
	\aap, 339, 737
\bibitem[Okumura et al.(2000)]{oku00} Okumura, S. K., et al.  2000,
	\pasj, 52, 393
\bibitem[Paczy\'nski(1998)]{pac98} Paczy\'nski, B.  1998, \apjl, 494, L45
\bibitem[Portegies Zwart et al.(1999)]{por99} Portegies Zwart, S. F.,
	Makino, J., McMillan, S. L. W., \& Hut, P.  1999, \aap, 348, 117
\bibitem[Ptak \& Griffiths(1999)]{pta99} Ptak, A., \& Griffiths, R. E.  1999,
	\apjl, 517, L85
\bibitem[Sakamoto(1999)]{sak99} Sakamoto, S.  1999, \apj, 523, 701
\bibitem[Sandage \& Tammann(1975)]{san75} Sandage, A., \& Tammann, G. A.  1975,
	\apj, 196, 313
\bibitem[Shen \& Lo(1995)]{she95} Shen, J., \& Lo, K. Y.  1995,
	\apjl, 445, L99
\bibitem[Shopbell \& Bland-Hawthorn(1998)]{sho98} Shopbell, P. L.,
	\& Bland-Hawthorn, J.  1998, \apj, 493, 129
\bibitem[Smith et al.(1991)]{smi91} Smith, P. A., Brand, P. W. J. L.,
	Mountain, C. M., Puxley, P. J., \& Nakai, N.  1991, \mnras, 252, 6p
\bibitem[Stevens et al.(1999)]{ste99} Stevens, I. R., Strickland, D. K.,
	\& Wills, K. A.  1999, \mnras, 308, L23
\bibitem[Sunada et al.(1994)]{sun94} Sunada, K., Kawabe, R.,
	\& Inatani, J.  1994, Int. J. Infrared Millimeter Waves, 14, 1251
\bibitem[Taniguchi et al.(2000)]{tan00} Taniguchi, Y., Shioya, Y.,
	Tsuru, T.G., \& Ikeuchi, S.  2000, \pasj, 52, 533
\bibitem[Tenorio-Tagle \& Bodenheimer(1988)]{ten88} Tenorio-Tagle, G.,
	\& Bodenheimer, P.  1988, \araa, 26, 145
\bibitem[Wei$\beta$ et al.(1999)]{wei99} Wei$\beta$, A., Walter, F.,
	Neininger, N., \& Klein, U.  1999, \aap, 345, L23
\bibitem[Wild et al.(1992)]{wil92} Wild, W., et al.  1992, \aap, 265, 447
\bibitem[Wills et al.(1999)]{wil99} Wills, K. A., Redman, M. P.,
	Muxlow, T. W. B., \& Pedlar, A.  1999, \mnras, 309, 395
\end{thebibliography}
\end{document}